\documentclass[conference]{IEEEtran}

\ifCLASSINFOpdf
\else
\fi
\usepackage{array}
\usepackage{mdwmath}
\usepackage{mdwtab}
\usepackage{eqparbox}

\usepackage{amsmath}
\usepackage{booktabs}
\usepackage{xcolor}
\usepackage{soul}
\usepackage{url}
\usepackage{multirow}
 \usepackage{graphicx}

\begin{document}
\title{PrepNet: A Convolutional Auto-Encoder to Homogenize CT Scans for Cross-Dataset Medical Image Analysis}

\author{\IEEEauthorblockN{Mohammadreza Amirian\IEEEauthorrefmark{1}\IEEEauthorrefmark{2},
Javier A. Montoya-Zegarra\IEEEauthorrefmark{1},
Jonathan Gruss\IEEEauthorrefmark{1}\IEEEauthorrefmark{5},
Yves D. Stebler\IEEEauthorrefmark{1}\IEEEauthorrefmark{5},\\
Ahmet Selman Bozkir\IEEEauthorrefmark{1}\IEEEauthorrefmark{5}, 
Marco Calandri\IEEEauthorrefmark{3},
Friedhelm Schwenker\IEEEauthorrefmark{2} and
Thilo Stadelmann\IEEEauthorrefmark{1}\IEEEauthorrefmark{4}}
\IEEEauthorblockA{\IEEEauthorrefmark{1}ZHAW School of Engineering, 8400 Winterthur, Switzerland}
\IEEEauthorblockA{\IEEEauthorrefmark{2}Ulm University, Institute of Neural Information Processing, 89081 Ulm, Germany}
\IEEEauthorblockA{\IEEEauthorrefmark{3}University of Turin, Department of Oncology, 10124 Turin, Italy}
\IEEEauthorblockA{\IEEEauthorrefmark{4}Fellow, ECLT European Centre for Living Technology, 30123 Venice, Italy}
\IEEEauthorblockA{\IEEEauthorrefmark{5}These authors contributed equally to this work}
\IEEEauthorblockA{amir@zhaw.ch}}
%\IEEEauthorblockA{\{amir, mony, bozk, stdm\}@zhaw.ch}, \{steblyve, grussjon\}@students.zhaw.ch, marco.calandri@unito.it, friedhelm.schwenker@uni-ulm.de}


\maketitle

\begin{abstract}
%\boldmath
With the spread of COVID-19 over the world, the need arose for fast and precise automatic triage mechanisms to decelerate the spread of the disease by reducing human efforts e.g. for image-based diagnosis. Although the literature has shown promising efforts in this direction, reported results do not consider the variability of CT scans acquired under varying circumstances, thus rendering resulting models unfit for use on data acquired using e.g. different scanner technologies. While COVID-19 diagnosis can now be done efficiently using PCR tests, this use case exemplifies the need for a methodology to overcome data variability issues in order to make medical image analysis models more widely applicable. In this paper, we explicitly address the variability issue using the example of COVID-19 diagnosis and propose a novel generative approach that aims at erasing the differences induced by e.g. the imaging technology while simultaneously introducing minimal changes to the CT scans through leveraging the idea of deep auto-encoders. The proposed prepossessing architecture (\emph{PrepNet}) ($i$) is jointly trained on multiple CT scan datasets and ($ii$) is capable of extracting improved discriminative features for improved diagnosis. Experimental results on three public datasets (SARS-COVID-2, UCSD COVID-CT, MosMed) show that our model improves cross-dataset generalization by up to $11.84$ percentage points despite a minor drop in within dataset performance.
\end{abstract}
~\\
\begin{keywords}
Adaptive preprocessing, domain adaptation, auto-encoder
\end{keywords}

% \fi
\IEEEpeerreviewmaketitle

%%%%%%%%%%%%%%%%%%%%%%%%%%%%%%%%%%%%%%%%%%%%%%%%%%%%%%%%%%%%%%%%%%%%%%%%%%%%%%%%%%%%%%%%%%%%%%%%
\section{Introduction}\label{sec:intro}
A major challenge in rolling out machine learned models to a broad user base is the variability of data encountered in the real world. Models can only be expected to work well on data of similar distribution as has been used for training, but ubiquitously, differences in image acquisition setup hinder the applicability of a once developed model in novel settings. A recent example for the negative effects of such failure to adapt between different domains has been given at the start of the COVID-19 pandemic:

As of 2\textsuperscript{nd} February 2021, this disease has caused over $100$ million infections worldwide and over $2$ million deaths according to the World Health Organisation (WHO)~\cite{who_reports}. To alleviate this, rapid diagnosis of COVID-19 cases has been proven to be effective for decelerating the spread of the disease~\cite{horry2020covid}. According to~\cite{horry2020covid,chen2020sars}, reverse transcriptase quantitative polymerase chain reaction (RT-qPCR) tests are accepted as the gold standard for the identification of positive cases. However, this type of test was not available in sufficient numbers at the beginning of the pandemic. Further, beyond being time-consuming, it relies on both human effort and expert knowledge. Thus, there arose a need for automatic diagnostic methods that can assist experts and reduce human efforts by targeting the automatic identification of COVID-19 positive cases. The literature has shown promising efforts in the automatic identification of COVID-19 cases from lung computed tomography (CT) scans using computer vision methods~\cite{mei2020nature,harmon2020nature,lin2020rnas,wang2020tmi}. 
Lessmann et al. addressed cross-vendor analysis (between different CT scanners such as Varian, Siemens, GE Healthcare, Philips and Canon) for 3D CT scans successfully \cite{lessmann2021automated}. However, it is demonstrated that a considerable drop in cross-dataset performance appears for the diagnosis of 2D CT scans acquired via different devices. Thus, the previously mentioned \textit{within dataset variability} has the potential to discourage the community to merge and annotate data from multiple sources. As a result, combining datasets is a challenge posed not only for COVID detection but also for other applications in diagnosis and segmentation.

In this paper, we address domain adaptation of medical image analysis methods by proposing a deep convolutional neural network (CNN) for preprocessing 2D CT scans such that it is trained to fool a classifier that discriminates between various CT datasets, thus aiming to remove the within dataset variability. We evaluate the performance of the suggested method on the exemplary use case of predicting COVID-19 positive cases, due to the global variability in respective datasets and the availability of plenty of opportunities to compare. It should be noted that, the methodology is inspired by generative adversarial learning \cite{DBLP:conf/nips/GoodfellowPMXWOCB14,schmidhuber2020generative}.
Our contribution is twofold: ($i$) we propose a novel trainable preprocessing CNN architecture with a dual training objective that is capable of equalizing the variability of different CT-scanner technologies in the image domain as a pre-processor (\emph{PrepNet}); ($ii$) we validate this model by showing the transferability of its diagnostic capabilities between different CT data sources based on common public benchmarks.
We conduct experiments on the \textit{SARS-CoV-2 CT-scan dataset}~\cite{DVN/SZDUQX_2020} and the \textit{UCSD COVID-CT dataset}~\cite{zhao2020covid} as well as \textit{MosMed dataset}~\cite{morozov2020mosmeddata}. Our results show that our \emph{PrepNet} model improves the cross-dataset COVID-19 diagnosis performance (i.e., training on one dataset and testing on another) by $11.84$ percentage points (pp) through creating a unified representation of multi-dataset CT scans.

%%%%%%%%%%%%%%%%%%%%%%%%%%%%%%%%%%%%%%%%%%%%%%%%%%%%%%%%%%%%%%%%%%%%%%%%%%%%%%%%%%%%%%%%%%%%%%%%
\section{Related Work}\label{sec:rel_work}
With the emergence of COVID-19, many studies and datasets have been proposed in the literature which show an increase in data diversity over time and the extent of related computer vision methods to deal with it \cite{cohen2020covid,gunraj2021covid}. Horry et al.~\cite{horry2020covid} utilize a transfer learning scheme to build various COVID-19 classifiers based on several off the shelf CNN models such as VGG16/19~\cite{simonyan2014very}, Resnet50~\cite{he2016deep}, InceptionV3~\cite{szegedy2016rethinking}, Xception~\cite{chollet2017xception}, and  InceptionResnet~\cite{szegedy2017inception}. They compared the generalization capability of various images sources such as X-ray, CT and ultrasound images and developed a pre-processing scheme for X-ray images to reduce noise at non-lung areas in order to decrease the effect of quality imbalance among the employed images. A VGG19~\cite{simonyan2014very} coupled with ultrasound images is found to yield the best validation accuracy of $99$\%, while $84$\% have been achieved using CT scans \cite{he2020sample}. 

He et al.~\cite{he2020sample} propose a sample-efficient learning concept called ``Self-Trans'' via synergetically combining transfer learning and contrastive self-supervised learning. They seek intrinsic visual patterns in CT scans without relying on labels created with human effort. Besides, they open-sourced their CT dataset involving $349$ COVID-19 positive patients and $397$ COVID-19 negatives \cite{zhao2020covid}. They achieve an accuracy of $86$\% through unbiased feature representations together with a reduction of overfitting.

Mobiny et al. \cite{mobiny2020radiologist} propose the DECAPS approach with following contributions: ($i$) inverted dynamic routing \cite{sabour2017dynamic} to avoid seeking visual features from non-related regions, ($ii$) training with a two-stage patch crop and drop strategy to encourage the network to focus on the useful areas, ($iii$) employing conditional generative adversarial networks for data augmentation. Experiments result $84.3$\% precision and $91.5$\% recall along with $87.6$\% accuracy. They additionally report results for the conventional deep classifiers DenseNet121 \cite{gao2017densely} and Resnet50 \cite{he2016deep}, yielding $82.5$\% and $80.8$\% accuracy, respectively. In contrast to this study, Pham \cite{pham2020comprehensive} points out the negative impact of data augmentation in the context of CT-based COVID-19 image classification. In his study, the author fine-tunes various well-known pre-trained CNN models ranging from AlexNet \cite{krizhevsky2012imagenet} to NasNet-Large \cite{zoph2018learning}. Experiments conducted on the already introduced CT dataset \cite{zhao2020covid} credit a DenseNet-201 with the best accuracy of $96.2$\%. However, data augmentation using random vertical/horizontal flips (p=$0.5$), vertical/horizontal translation ($\pm 30$ pixels) and scaling ($\pm 10$\%)  yields a $6$\% accuracy drop on average.

Chaganti et al. \cite{chaganti2020quantification} suggest a deep-reinforcement-learning-based scheme focusing on seeking doubtful lung areas on CT scans to localize abnormal portions. A recent study by \cite{gunraj2021covid}, a novel architecture called ``COVID-Net- CT-2'' which utilizes machine-driven design exploration based on iterative constrained optimization is proposed \cite{wong2019netscore}. The authors point out that one of the subtle problems of earlier studies is the limited number of patients and poor diversity of CT scans in terms of multi-nationality. Therefore, they introduce the two large-scale COVID-19 CT datasets called ``COVIDx CT-2A'' and ``COVIDx CT-2B'' gathered from $4,501$ patients from at least $15$ countries, totally comprising $194.922$ and $201.103$ images respectively. Experiments show that the architecture achieves a sensitivity of $99.0$\% and an accuracy of $98.1$\%, which competes with radiologist-level decision making capability.
The study deals with variability in the patients' ethnicity, while CT scans generated by various vendors' devices exhibit visual differences, artifacts, and variable intensities that are never addressed so far. Thus, independent from the reported success of some deep learning architecture, it is likely to witness a drop in prediction accuracy during inference when a test image is taken with a different device as has been used for training. Motivated by this issue, we propose to employ a pre-processing network (\emph{PrepNet}) to standardize CT images with respect to the visual differences among datasets prior to training of any final diagnosis model, relying on generative architectures since they showed very promising results for similar tasks \cite{mobiny2020radiologist}. An advantage of this approach is that the \emph{PrepNet} can be combined with any downstream diagnosis model, thus leveraging future progress there without additional costs while improving cross-dataset performance.

Two research papers closely related to the goal of domain adaptation in this study are presented by Lessmann et al. addressing cross-vendor diagnosis \cite{lessmann2021automated} and Amyar et al. using auto-encoders in multi-task learning \cite{amyar2020multi}. Nevertheless, Lessmann et al. did not confront a considerable cross-vendor performance drop because of using a richer source of information (3D scans) as explained in \cite{de2020improving}. Amyar et al. leveraged multi-task learning and trained an auto-encoder besides a segmentation and classification model for COVID-19 diagnosis. However, they did not aim at removing the cross-dataset variability of the scans. This study focuses on homogenizing the 2D CT scans by reducing cross-dataset information.  

\begin{figure*}[t!]
    \centering
    \includegraphics[width=0.9\linewidth]{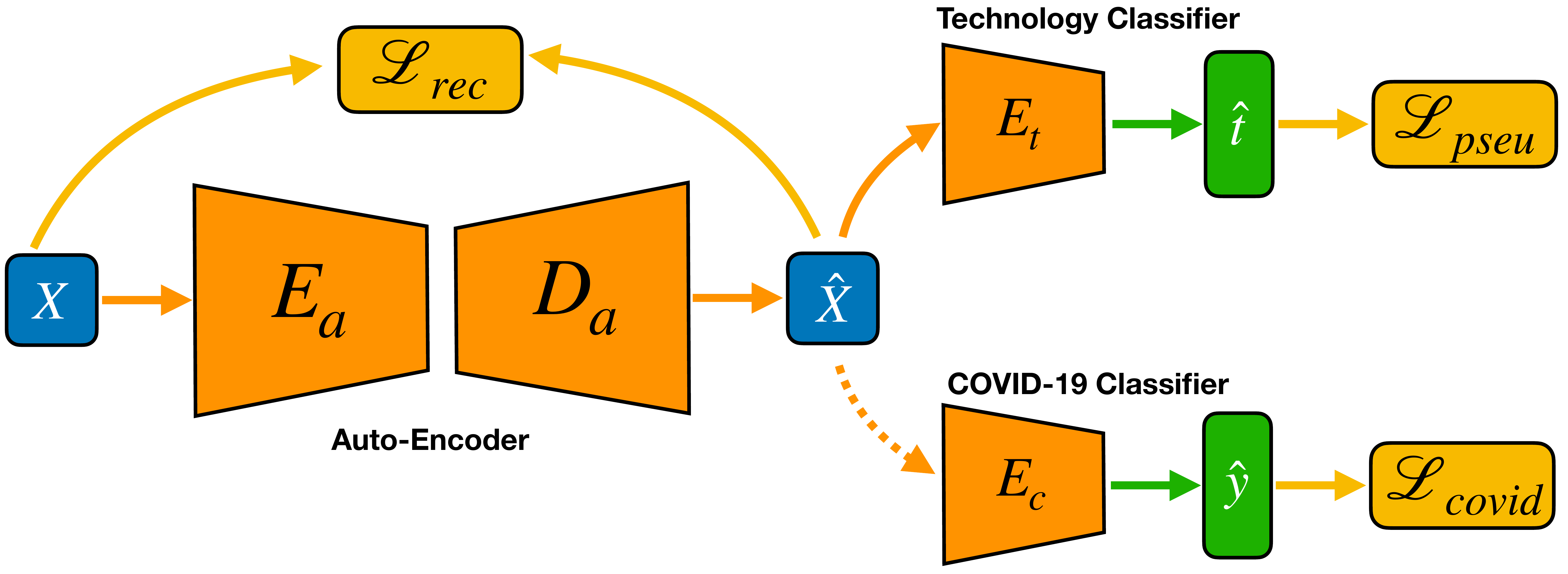}
    \caption{The architecture of our proposed \textit{PrepNet} model consists of three main modules: ($i$) an auto-encoder that acts as a CT cross-dataset homogenizer; ($ii$) a multi CT-technology classifier; and ($iii$) a COVID-19 binary classifier. The auto-encoder and the multi CT technology classifier are trained adversarially. The binary COVID-19 classifier is independently trained using the auto-encoder's output.}
    \label{fig:architecture}
\end{figure*}

\section{Methodology}\label{sec:method}
In this section, we give details of our \textit{PrepNet} model in terms of network architecture, core modules, and loss functions. 
%\subsection{Method Formulation}\label{sec:formulation}
The architecture of our proposed model is presented in Figure~\ref{fig:architecture}.
For a group of $\mathcal{N}$ input CT scans $\{\mathcal{X}^{n}\}^{\mathcal{N}}_{n=1}$, coming from different CT vendors' devices, our model extracts multi-scale discriminative feature maps through an auto-encoder and reconstructs the original CT scans $\{\mathcal{\hat{X}}^{n}\}^{\mathcal{N}}_{n=1}$. The reconstructed CT scans are next fed into a dataset/technology classification branch which acts as a pseudo-label classifier and is responsible for discriminating among different CT datasets. Once this model is trained end-to-end in an adversarial way, the reconstructed CT scans are fed into a COVID-19 classifier which is trained directly on the reconstructed CT-scans. The COVID-19 classification branch is responsible for the classification of healthy vs. non-healthy patients. The complete network model with its main modules are described in more detail below.

\subsection{Model Architecture}\label{sec:network_model}
\noindent\textbf{Auto-Encoder Module:}  We feed a CT scan image $\mathcal{X}^{n}$ into our auto-encoder ($E_{a}$ and $D_{a}$) and obtain a reconstructed version $\mathcal{\hat{X}}^{n}$ given by $\mathcal{\hat{X}}^{n}=D_{a}(E_{a}(\mathcal{X}^{n}))$. The encoder $E_{a}$ is based on the standard classification network VGG-Net~\cite{simonyan2014very}, whilst the decoder $D_{a}$ is a convolutional network with the same number of layers as the encoder. We add skip-connections from $E_{a}$ to $D_{a}$ to recover the spatial information lost during the down-sampling operations.

\noindent\textbf{Dataset Classifier Module:} The CT dataset classifier $E_t$ receives the reconstructed CT scan $\mathcal{\hat{X}}^{n}$ from the auto-encoder as input and feeds it into an encoder branch $E_t(\mathcal{\hat{X}}^{n})$ that classifies the CT dataset/technology. In our experiments, $E_t$ relies on the VGG-Net architecture as well.

\noindent\textbf{COVID-19 Classifier Module:} The COVID-19 classifier $E_c$ is also uses several backbone architectures. Given a reconstructed CT scan $\mathcal{\hat{X}}^{n}$, it outputs COVID vs. non-COVID predictions, i.e. $E_c(\mathcal{\hat{X}}^{n})$.

\subsection{Loss Functions and Evaluation Metric}\label{sec:loss_functions}
The complete loss function of \textit{PrepNet} is based on the various terms presented in Figure~\ref{fig:architecture}. It comprises a reconstruction loss $\mathcal{L}_{rec}$ and two classification losses $\mathcal{L}_{pseu}$ and $\mathcal{L}_{covid}$:
\begin{equation}
    \mathcal{L}_{total} = \mathcal{L}_{rec} + \mathcal{L}_{pseu} + \mathcal{L}_{covid}
\end{equation}
Given the labeled dataset $\mathcal{D}=\{(\mathcal{X}^{n}, y^{n}, p^{n})\}_{n}^{N}$ comprising the CT scans $\mathcal{X}^{n}$ together with their binary COVID label $y^{n}$ and the CT-dataset pseudo label $p^{n}$, the auto-encoder reconstruction loss is given by $\mathcal{L}_{rec}=\sum_{n}\|\mathcal{X}^{n}-\mathcal{\hat{X}}^{n} \|_ {2}^{2}$; the COVID-19 binary classification loss is denoted $\mathcal{L}_{covid}=-\sum_{n}y_{n}\log \hat{y}_{n} + (1-y_{n})\log(1-\hat{y}_{n})$; the CT dataset pseudo label is computed by $\mathcal{L}_{pseu}=-\sum_{n}p_{n}\log \hat{p}_{n}$.

To measure the COVID-19 detection performance and to minimize the effect of class imbalance in datasets, we use the balanced accuracy metric (BA)~\cite{brodersen2010balanced}
\begin{equation}
    BA~=~\frac{TP}{P} + \frac{TN}{N}
\end{equation}
where $P$ and $N$ are the number of positive and negative samples respectively and $TP$ and $TN$ denote the number of true positive and true negative predictions, respectively. In addition, we also use specificity, sensitivity, and area under the curve to evaluate the COVID-19 performance results.

%%%%%%%%%%%%%%%%%%%%%%%%%%%%%%%%%%%%%%%%%%%%%%%%%%%%%%%%%%%%%%%%%%%%%%%%%%%%%%%%%%%%%%%%%%%%%%%%
\section{Experiments}\label{sec:experiments}

\subsection{Datasets}\label{subsec:datasets}

We use three public datasets to validate our approach experimentally. The \textit{SARS-CoV-2 CT-scan dataset}~\cite{DVN/SZDUQX_2020} comprises a total of $4,173$ CT images of real patients from the Public Hospital of the Government Employees of Sao Paulo (HSPM) and the Metropolitan Hospital of Lapa, both in Sao Paulo - Brazil ($2,168$ positive/infected and $768$ healthy patients). Moreover, $1,247$ CT scans belong to patients who have other pulmonary diseases. The CT image annotations (positive vs. negative) have been done by three different clinicians. Note that during our visual inspection we found two erroneous images (i.e. unrelated to the problem domain) and excluded them from the dataset. In addition, we also excluded the $1,247$ pulmonary diseased patients. 

The \textit{UCSD COVID-CT dataset}~\cite{zhao2020covid} has been collected in the Tongji Hospital in Wuhan, China during the outbreak of COVID-19 between the months of January/2020 and April/2020. This dataset contains $349$ CT images from infected patients and $397$ from non-infected patients. All images have been annotated by a senior radiologist of the same hospital. As reported by \cite{mobiny2020radiologist}, heights of the images in this dataset range between $153$ and $1,853$ pixels with an average of $491$ pixels, whereas the widths vary between $124$ and $1,458$ pixels (average of $383$ pixels). For partitioning, we follow the splitting guideline provided by the authors of the dataset. Table \ref{tab:datasets} summarizes the train, validation and test splits for each dataset.

\begin{figure}[ht]
    \centering
    \resizebox{\linewidth}{!}{\centering
\begin{tabular}{l | c c c c}
    \toprule
    COVID & SARS-COV-2 & UCSD COVID-CT & MosMed COVID-19 \\
    \toprule
    Negative &
    \includegraphics[width=3cm]{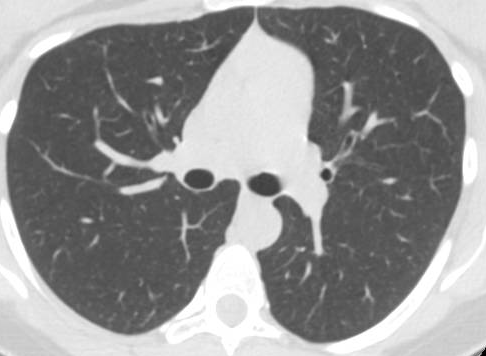} &
    \includegraphics[width=3cm]{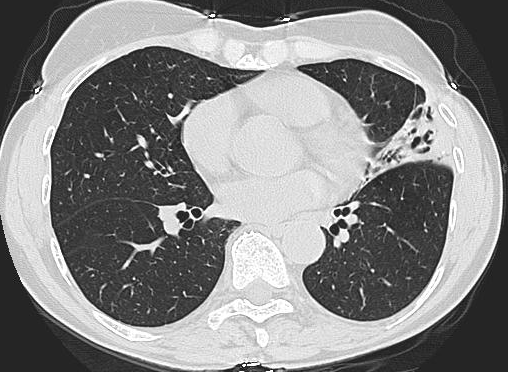} &
    \includegraphics[width=3cm]{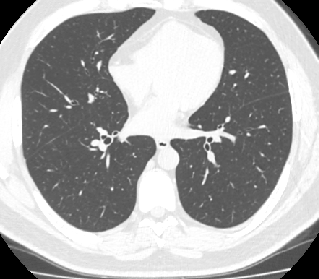} \\
    \toprule
    Positive &
    \includegraphics[width=3cm]{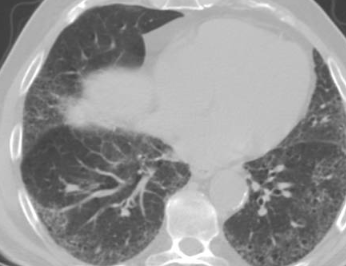} &
    \includegraphics[width=3cm]{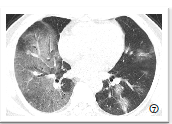} &
    \includegraphics[width=3cm]{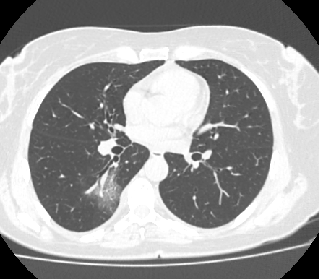} \\
    \toprule     
\end{tabular}}
    %\caption{A sample of each dataset that is classified as COVID-19 positive and one that is classified as COVID-19 negative. Transformations to the MosMed COVID-19 dataset~\cite{morozov2020mosmeddata} that rotate and crop the CT scans have been applied already.}
    \caption{COVID-19 positive and negative samples for each used dataset. Note the variabilities in terms of texture, size, and shape across datasets.}
    \label{fig:figure_exp2_ct}
\end{figure}

The \textit{MosMed dataset}~\cite{morozov2020mosmeddata} was collected by the Moscow Health Care Department from different municipal hospitals in Russia between March/2020 and April/2020. The dataset contains axial CT images from $1110$ patients with different levels of COVID-19 severity, ranging from mild to critical cases and also healthy patients. Some image samples of each dataset are provided in Figure~\ref{fig:figure_exp2_ct}.

\begin{table*}[t!]
    \begin{center}
    \resizebox{0.75\textwidth}{!}{  
        \begin{tabular}{l|ccc|ccc}
            \toprule
             &  &  &  & \multicolumn{3}{c}{Dataset portions} \\
             Dataset  & Type & Size & Country & Train & Validation & Test \\
            \midrule
            SARS-COV-2~\cite{DVN/SZDUQX_2020} & 2D CT & Various & Brazil & $2,046$ ($70$\%) & $439$ ($15$\%)& $439$ ($15$\%) \\
            UCSD COVID-CT~\cite{zhao2020covid} & 2D CT & Various & China & $423$ ($57$\%) & $116$ ($16$\%) & $201$ ($27$\%) \\
            MosMed Dataset~\cite{morozov2020mosmeddata} & 3D CT & Various & Russia & \multicolumn{3}{c}{$1100$ images for unseen test dataset} \\
            \toprule
        \end{tabular}
    }
    \caption{Public datasets used in our study together with their corresponding data splits. The SARS-COV-2~\cite{DVN/SZDUQX_2020} and the UCSD COVID-CT~\cite{zhao2020covid} datasets are used for training and evaluating our models, while the MosMed dataset~\cite{morozov2020mosmeddata} is used for evaluation purposes only.}
    \label{tab:datasets}
    \end{center}
\end{table*}

%\hl{add illustrative images here from both datasets. ideally a table with two rows. first row dataset A, second row dataset B. along the columns we put positive and negative covid smaples. the details on the images should be ideally clearly seen.}

\subsection{Implementation Details}\label{subsec:imp_details}
We run all our experiments using the publicly available Pytorch 1.5.0 library and an NVIDIA VP100 GPU ($32$ GB of VRAM). During network training, each image is first resized according to the input size of the classifiers' backbones; we use histogram equalization as a fixed preprocessing step, then apply the mean and standard deviation of ImageNet pretrained models. We train \emph{PrepNet} using the AdamW optimizer \cite{loshchilov2017decoupled}. We perform a $24$ hour hyperparameter search with six parallel runs using the Bayesian search strategy with Hyperband for early-stopping on one GPUs \cite{tuggener2019automated}. The hyperparameter search improves the chance of avoiding local minima and presenting optimal results of every configuration. The best model is selected based on the optimal validation performances. During training, we first train the auto-encoder for $20$ epochs and warm up the dataset classification branch for $2$ epochs before we start with the adversarial training. Once the adversarial training is finished, we train the COVID classification branch independently from the other two branches using the output of the auto-encoder/\emph{PrepNet}.

\subsection{Experimental Results}\label{subsec:results}
The within- and cross-dataset performance of the proposed preprocessing schemes are presented in Table~\ref{tab:covid_baselines}. In order to observe possible overfitting, we report the hold out test set performance on each dataset. The cross-dataset performance is evaluated by measuring the balanced accuracy (minimizing the effect of class imbalance) of the models trained on one dataset and tested on the other. We report results using the balanced accuracy of the models trained on the SARS-COV-2 and UCSD COVID-CT datasets. Further metrics also include sensitivity (Sens), specificity (Spec) and area under the curve (AUC). In the rows, we present the datasets used during training. Furthermore, we group the results by model. The first group of results are related to the COVID classifier (VGG-19 pre-trained model), that is trained and evaluated on the original CT scans. The second group of results is related to the auto-encoder alone trained on both datasets in a self-supervised manner to minimize the reconstruction loss. The third group of results relate to full \emph{PrepNet} preprocessing before training the classifiers.
\begin{table*}
\centering
\resizebox{1.0\textwidth}{!}{
    \begin{tabular}{l|cccc|cccc|c|c|c}
        \toprule
        Test dataset $\rightarrow$ & \multicolumn{4}{c|}{SARS-COV-2} & \multicolumn{4}{c|}{UCSD COVID-CT} & Within Test & Cross-Dataset & Pre-trained  \\ 
        Dataset portion  & BA  & Sens     & Spec     & AUC      & Test  & Sens      & Spec     & AUC    & Average & Average & encoder\\
        \midrule
        \midrule
        & \multicolumn{11}{c}{\emph{COVID classifier}} \\
        \midrule
        SARS-COV-2      & $0.8924$  & $0.9292$ & $0.7876$ & $0.8584$ & $0.4433$ & $0.7835$  & $\boldsymbol{0.1262}$ & $0.4548$ & $\boldsymbol{0.8587}$ & $0.4159$ & \multirow{2}*{Yes}\\
        UCSD COVID-CT   & $0.3295$ & $0.3476$ & $0.2743$ & $0.3110$ & $\boldsymbol{0.8250}$  & $0.7113$  & $\boldsymbol{0.9320}$ & $\boldsymbol{0.8216}$ & (baseline) & (baseline) & \\ 
        \midrule
        \midrule
        & \multicolumn{11}{c}{\emph{AutoEncoder}} \\
        \midrule
        SARS-COV-2      & $0.8956$ & $\boldsymbol{0.9907}$ & $0.6460$ & $0.8183$ & $0.4983$ & $0.9175$   & $0.0970$ & $0.5073$   & $0.8555$ & $0.4836$ & \multirow{2}*{Yes}\\
        UCSD COVID-CT   & $0.49405$ & $0.6030$ & $\boldsymbol{0.3008}$ & $0.4519$ & $0.8154$  & $0.7216$ & $0.8846$ & $0.8031$ & ($-0.32\%$) & ($+6.77\%$) & \\
        \midrule
        \midrule
        & \multicolumn{11}{c}{\emph{PrepNet}} \\
        \midrule
         SARS-COV-2      & $\boldsymbol{0.9007}$ & $0.9353$ & $\boldsymbol{0.7982}$ & $\boldsymbol{0.8668}$ & $\boldsymbol{0.5157}$ & $\boldsymbol{0.9175}$  & $0.1067$ & $\boldsymbol{0.5121}$ & $0.8404$ & $\boldsymbol{0.5343}$ & \multirow{2}*{Yes}\\
        UCSD COVID-CT   & $\boldsymbol{0.5545}$ & $\boldsymbol{0.6446}$ & $0.1858$ & $\boldsymbol{0.4852}$ & $0.7800$ & $\boldsymbol{0.8556}$   & $0.7087$ & $0.7822$  & ($-1.83\%$) & ($+11.84\%$) & \\
        \toprule
    \end{tabular}}
    \caption{Test performance of different baselines compared to our \emph{PrepNet} model. Results demonstrate that our model is capable of increasing the cross-dataset average.}
    \label{tab:covid_baselines}
\end{table*}

The results in Table~\ref{tab:covid_baselines} show that the average cross-dataset performance (over all dataset splits) of models trained on original data increases by $6.77$pp after using the pure auto-encoder model, and by $11.84$pp through \emph{PrepNet}. However, the average test accuracy for within-dataset evaluation declines by $0.32$pp and $1.83$pp after applying the baseline auto-encoder or \emph{PrepNet}, respectively. A discussion regarding this effect is presented in the next section.

In our experiments, we use the VGG19~\cite{simonyan2014very} as the baseline model because it is more straight-forward to train and has shown good generalization properties on 2D medical images based on previous practical experiments\footnote{\url{https://stanfordmlgroup.github.io/competitions/mura/}}. Besides that, the VGG architecture has been also successfully applied for COVID-19 identification~\cite{horry2020covid, he2020sample}. 

As part of our ablation study, we also evaluated how different backbones affect the COVID-19 diagnosis accuracy of \emph{PrepNet}. More precisely, we replicate the experiments for each dataset (SARS-COV-2 and UCSD COVID-CT) and evaluate different CNN architectures as part of our COVID-classifier Module (See Section \ref{sec:network_model} for more information). The CNN architectures include ResNet18~\cite{he2016deep}, Inception~\cite{szegedy_inception}, and EfficientNet-B0~\cite{tan2019efficientnet}. We report results in  Table~\ref{tab:prepnet_backbones}. Experimental results show that in almost all backbones, the average cross-dataset performance increases with the cost of a small decrease in the within-dataset accuracy.
\begin{table*}
\centering
\resizebox{1.0\textwidth}{!}{
    \begin{tabular}{l|cccc|cccc|c|c|c}
        \toprule
        Test dataset $\rightarrow$ & \multicolumn{4}{c|}{SARS-COV-2} & \multicolumn{4}{c|}{UCSD COVID-CT} & Within Test & Cross-Dataset & Pre-trained  \\ 
        Dataset portion  & BA & Sens     & Spec     & AUC      & Test   & Sens      & Spec     & AUC    & Average & Average & encoder\\
        \midrule
        \midrule
        & \multicolumn{11}{c}{\emph{VGG19}} \\
        \midrule
        SARS-COV-2      & $\boldsymbol{0.9007}$ & $\boldsymbol{0.9353}$ & $0.7982$ & $\boldsymbol{0.8668}$ & $0.5157$ & $0.9175$ & $0.1067$ & $0.5121$ & $\boldsymbol{0.8404}$ & $\boldsymbol{0.5343}$ & \multirow{2}*{Yes}\\
        UCSD COVID-CT   & $\boldsymbol{0.5545}$ & $\boldsymbol{0.6446}$ & $0.1858$ & $\boldsymbol{0.4852}$ & $0.7800$ & $0.8556$ & $0.7087$ & $0.7822$ & ($-1.83\%$) & ($+11.84\%$) & \\
        \midrule
        \midrule
        & \multicolumn{11}{c}{\emph{ResNet18}} \\
        \midrule
        SARS-COV-2      &  $0.7462$   & $0.7046$ & $\boldsymbol{0.8584}$ & $0.7815$ & $0.4728$  & $0.8144$ & $0.1538$ & $0.4841$ & $0.7345$ & $0.4940$ & \multirow{2}*{Yes}\\
        UCSD COVID-CT   & $0.5152$  & $0.6246$ & $0.1947$ & $0.4096$ & $0.7228$  & $0.8351$ & $0.6154$ & $0.7252$  & ($-12.42\%$) & ($+7.81\%$) & \\
        %\midrule
        %\midrule
        %& \multicolumn{11}{c}{\emph{ResNet50}} \\
        %\midrule
        %SARS-COV-2     & $0.8007$ & $0.8123$ & $0.76106$ & $0.7867$ & $\boldsymbol{0.5371}$ & $0.5464$ & $\boldsymbol{0.5288}$ & $\boldsymbol{0.5376}$ & $0.8057$ & $0.3727$ & \multirow{2}*{Yes}\\
        %UCSD COVID-CT  & $0.2083$ & $0.1908$ & $0.2478$ & $0.2193$ & $\boldsymbol{0.8168}$  & $\boldsymbol{0.8557}$ & $\boldsymbol{0.77885}$ & $\boldsymbol{0.8172}$ & ($-5.3\%$) & ($-4.32\%$) & \\
        \midrule
        \midrule
        & \multicolumn{11}{c}{\emph{Inception}} \\
        \midrule
        SARS-COV-2      & $0.8553$  & $0.9046$ & $0.7080$ & $0.8063$ & $0.4703$  & $\boldsymbol{0.9485}$ & $0.02885$ & $0.4886$ & $0.8286$ & $0.3995$ & \multirow{2}*{Yes}\\
        UCSD COVID-CT   & $0.3288$  & $0.36308$ & $0.2212$ & $0.2922$ & $0.8020$ & $0.8351$ & $0.7692$ & $0.8021$  & ($-3.01\%$) & ($-1.64\%$) & \\
                \midrule
        \midrule
        & \multicolumn{11}{c}{\emph{EfficientNet-B0}} \\
        \midrule
        SARS-COV-2      & $0.8735$  & $0.8923$ & $0.8142$ & $0.8532$ & $0.5223$  & $0.5979$ & $0.4519$ & $0.5249$ & $0.8253$ & $0.4835$ & \multirow{2}*{Yes}\\
        UCSD COVID-CT   & $0.4447$  & $0.5015$ & $\boldsymbol{0.2743}$ & $0.3879$ & $0.7772$  & $0.8041$ &  $0.7500$ & $0.7771$ & ($-3.34\%$) & ($+6.76\%$) & \\
        \toprule
    \end{tabular}}
    \caption{Experimental results of \emph{PrepNet} with different backbones: VGG19~\cite{simonyan2014very}, ResNet18~\cite{he2016deep}, Inception~\cite{szegedy_inception}, and EfficientNet-B0~\cite{tan2019efficientnet}. Note that \emph{PrepNet} increases the cross-dataset average.}
    \label{tab:prepnet_backbones}
\end{table*}

Finally, in order to evaluate the generalisation capabilities of \emph{PrepNet} and our baselines, we evaluate how our trained models perform on an unseen dataset, i.e. the \textit{MosMed dataset}~\cite{morozov2020mosmeddata}. The results in Table~\ref{tab:mosmed} show the improvements of our \emph{AutoEncoder} and \emph{PrepNet} models in terms of BA and sensitivity, however, with a decrease in specificity and AUC when compared with the COVID-19 classifier. Despite the decrease in specificity, we argue that especially for medical diagnosis and screening, a low specificity is less harmful than a reduction in sensitivity, as false positive cases can be discarded by additional examinations. On the contrary, a higher sensitivity is important as false negatives should be low.
\begin{table}
\centering
\resizebox{0.5\textwidth}{!}{
    \begin{tabular}{l|cccc|c}
        \toprule
        Test dataset $\rightarrow$ & \multicolumn{4}{c|}{MosMed} &   Pre-trained  \\ 
        Preprocessing  & BA & Sens     & Spec     & AUC & encoder\\
        \midrule
        \multirow{2}*{\emph{COVID-classifier}}      & $0.6066$   & $0.5246$ & $\boldsymbol{0.8771}$ & $\boldsymbol{0.7009}$ &    \multirow{2}*{Yes}\\
        & (baseline)    & (baseline) &  (baseline) & (baseline) & \\ 
        \midrule
        \multirow{2}*{\emph{AutoEncoder}}      & $0.6693$     & $0.7142$ & $0.5175$ & $0.6159$ &   \multirow{2}*{Yes}\\
                        &  ($+6.27\%$) & ($+18.96\%$) & ($-35.96\%$) & ($-8.50\%$) \\
        \midrule
        \multirow{2}*{\emph{PrepNet}}      & $\boldsymbol{0.7073}$ & $\boldsymbol{0.7558}$ & $0.5438$ & $0.6498$ &  \multirow{2}*{Yes}\\
                        & ($+10.07\%$) & ($+23.12\%$) & ($-33.33\%$) & ($-5.11\%$) \\
        \toprule
    \end{tabular}}
    \caption{Experimental COVID-19 classification results of the trained COVID-19 classifier, Auto-Encoder, and \emph{PrepNet} models on the Mosmed~\cite{morozov2020mosmeddata} unseen dataset.}
    \label{tab:mosmed}
\end{table}

\subsection{Discussion}\label{sec:discussion}
\begin{figure*}[ht]
     \centering
     \resizebox{0.6\textwidth}{!}{\centering
\begin{tabular}{l | c c c}
    \toprule
    Dataset & Original & Baseline auto-encoder & \emph{PrepNet} \\
    \toprule
    SARS-COV-2 &
    \includegraphics[width=3.5cm]{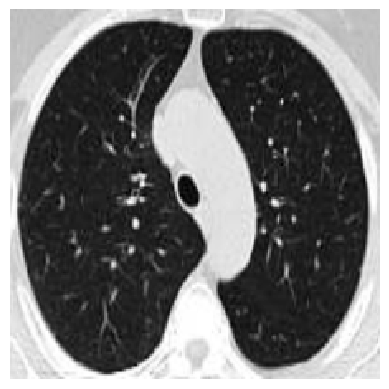} &
    \includegraphics[width=3.5cm]{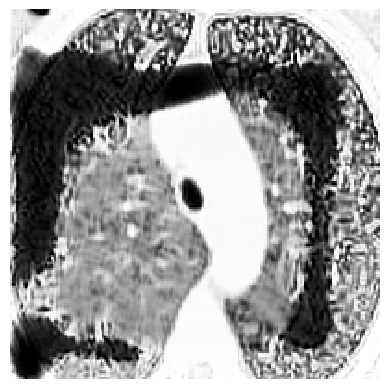} &
    \includegraphics[width=3.5cm]{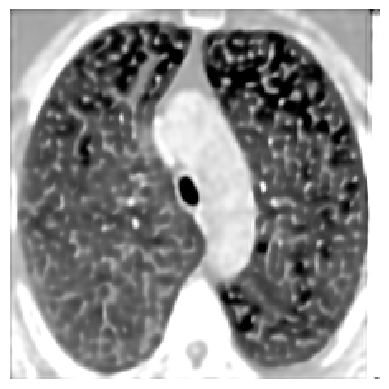} \\
    \toprule
    UCSD COVID-CT &
    \includegraphics[width=3.5cm]{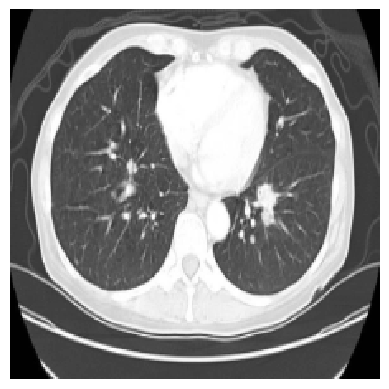} &
    \includegraphics[width=3.5cm]{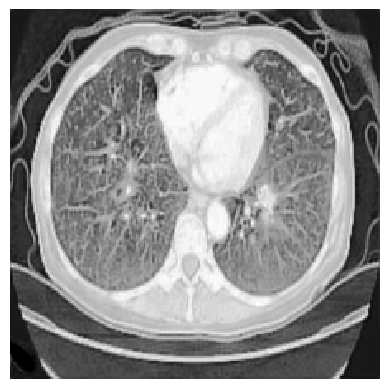} &
    \includegraphics[width=3.5cm]{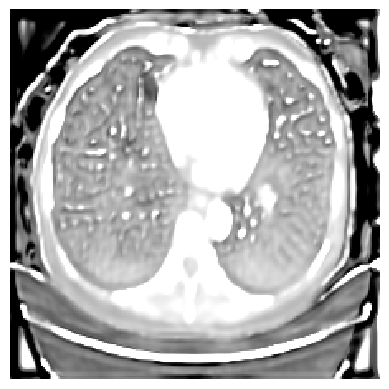} \\
    \toprule     
\end{tabular}}
     \caption{Severe cases of artifacts generated by the baseline and the proposed \emph{PrepNet}. The images demonstrate different levels of distortions like e.g. extreme contrasts.}
     \label{fig:artifacts}
\end{figure*}

\begin{figure}[ht]
    \centering
    \resizebox{\linewidth}{!}{\centering
\begin{tabular}{l | c c c c}
    \toprule
    Dataset & pre-processed & initial reproduction & \emph{PrepNet} reproduction \\
    \toprule
    SARS-COV-2 &
    \includegraphics[width=3.5cm]{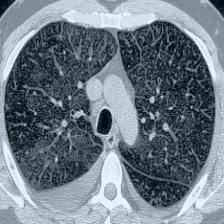} &
    \includegraphics[width=3.5cm]{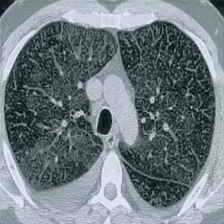} &
    \includegraphics[width=3.5cm]{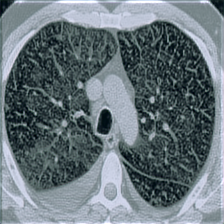} \\
    \toprule
    UCSD COVID-CT &
    \includegraphics[width=3.5cm]{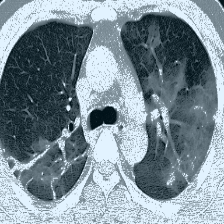} &
    \includegraphics[width=3.5cm]{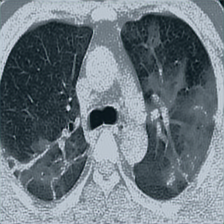} &
    \includegraphics[width=3.5cm]{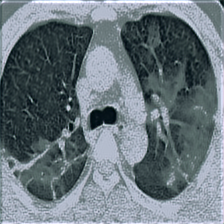} \\
    \toprule     
\end{tabular}}
    \caption{Samples CT scans that are wrongly classified after the trainable preprocessing.}
    \label{fig:missclassfy}
\end{figure}

The baseline and proposed pre-processing approaches introduce performance drops when applied before within-dataset classification. These approaches usually reduce the test accuracies when trained and evaluated on the same dataset using the corresponding dataset splits. Therefore, we further investigate the intermediate results of the baseline auto-encoder and \emph{PrepNet} on a case-by-case basis. 
Severe cases of generated artifacts through reconstruction via the baseline auto-encoder and the \emph{PrepNet} are presented in Figure \ref{fig:artifacts}. We conjecture that the drop in within-dataset test performance is caused by occasional artifacts such as these. These quality drops can be clearly seen in the reconstruction loss. However, it is not straightforward to correct them.
We could eventually overcome this by also investigating different data-augmentation strategies and by improving the network architecture of our auto-encoder. Additionally, we depict sample images in which the models failed to make a correct decision after auto-encoder or \emph{PrepNet} (See Fig. ~\ref{fig:missclassfy}). 
Limited amount of training data and noisy labels of public datasets are other factors contributing to low classification accuracies. One possible way to tackle this limitation is to rely on weakly supervised learning methods to improve the COVID-19 classification accuracy with the methodology summarized in ~\cite{simmler2021survey}. 

%%%%%%%%%%%%%%%%%%%%%%%%%%%%%%%%%%%%%%%%%%%%%%%%%%%%%%%%%%%%%%%%%%%%%%%%%%%%%%%%%%%%%%%%%%%%%%%%
\section{Conclusions and Future Work}\label{sec:conclusions}
In this paper, we introduced a novel approach to unify several CT scan datasets with respect to varying image datasets and acquisition circumstances such as CT scanner technology through training an adaptive pre-processing network that removes such specificities from the images themselves. Additionally, we presented initial results demonstrating the applicability of the method on three publicly available benchmark datasets. This way, it is possible to shift the focus of model training from merely optimizing hold-out test set performance on \emph{the same} data distribution (which likely does not transfer to any other environment) towards cross-dataset detection accuracy. The proposed \emph{PrepNet} improves the cross-dataset balanced accuracy by a margin of $11.84$ percentage points (\textit{SARS-CoV-2 CT-scan dataset}~\cite{DVN/SZDUQX_2020}) at the expanse of a decline in the within dataset test performance of ca. $1.83$pp  (\textit{UCSD COVID-CT database}~\cite{zhao2020covid}). These results suggest that the trainable preprocessing network erases some of the necessary information for diagnosis, due to artifacts. This information could be partially retained by propagating the gradients of the COVID-19 classifier network through the preprocessing model, and generated artifacts could be detected automatically by monitoring the reconstruction loss of the auto-encoder module. This, together with further investigations on the applicability and generality of the proposed approach to combine multiple datasets, is an intriguing theme for future research.

% final results
%\input{cisp_bmei/tables2/table_covid_vgg19}
%\input{cisp_bmei/tables2/backbones2}
%\input{cisp_bmei/tables2/mosmed_eval}

%\pagebreak
%\input{cisp_bmei/tables2/table_covid_vgg19_2}
%\input{cisp_bmei/tables2/backbones2}
%\input{cisp_bmei/tables2/mosmed_eval2}

\section*{Acknowledgment}
This research was financially supported by the ZHAW Digital Futures Fund under contracts ``\emph{SDMCT---Standardized Data and Modeling for AI-based CoVID-19 Diagnosis Support on CT Scans}'' as well as ``\emph{Synthetic data generation of CoVID-19 CT/X-rays images for enabling fast triage of healthy vs. unhealthy patients}''.

%%%%%%%%%%%%%%%%%%%%%%%%%%%%%%%%%%%%%%%%%%%%%%%%%%%%%%%%%%%%%%%%%%%%%%%%%%%%%%%%%%%%%%%%%%%%%%%%
\bibliographystyle{IEEEtran}
\bibliography{main.bib}

\end{document}